# Hybrid Functional Study Rationalizes the Simple Cubic Phase of Calcium at High Pressures

Hanyu Liu, Wenwen Cui, and Yanming Ma*

*State Key Laboratory of Superhard Materials, Jilin University, Changchun 130012, China*

Simple cubic (SC) phase has been long experimentally determined as the high-pressure phase III of elemental calcium (Ca) since 1984. However, recent density functional calculations within semi-local approximation showed that this SC phase is structurally unstable by exhibiting severely imaginary phonons, and is energetically unstable with respect to a theoretical body-centered tetragonal $I4_1/amd$ structure over the pressure range of phase III. These calculations generated extensive debates on the validity of SC phase. Here we have re-examined the SC structure by performing more precise density functional calculations within hybrid functionals of Heyd-Scuseria-Erhzerhof (HSE) and PBE0. Our calculations were able to rationalize fundamentally the phase stability of SC structure over all other known phases by evidence of its actual energetic stability above 33 GPa and its intrinsically dynamical stability without showing any imaginary phonons in the entire pressure range studied. We further established that the long-thought theoretical $I4_1/amd$ structure remains stable in a narrow pressure range before entering SC phase and is actually the structure of experimental Ca-III' synthesized recently at low temperature 14 K as supported by the excellent agreement between our simulated X-ray diffraction patterns and the experimental data. Our results shed strong light on the crucial role played by the precise electron exchange energy in a proper description of the potential energy of Ca.

High-pressure polymorphs of elemental Ca exhibit complicated and somewhat mysterious behaviors. X-ray diffraction (XRD) experiments established that at room temperature Ca transforms from its ambient-pressure face-centered cubic structure (FCC, Ca-I) to the body-centered cubic structure (BCC, Ca-II) at 19 GPa,[1] and then to a simple cubic structure (SC, Ca-III) at 32 GPa.[2-5] This transition sequence is rather unusual as accompanied by a decrease of coordination numbers (12 → 8 → 6). Ca-III remains remarkably stable up to 113 GPa and then transforms to Ca-IV [3]. Several other higher pressure phases of Ca-V, Ca-VI and Ca-VII were also observed at 139, 158 and 210 GPa, respectively.[3, 4, 6, 7] One intriguing physical phenomenon of Ca is the observed superconductivity[8] with $T_c$ = 2 K at 44 GPa as the consequence of phase transitions. It is strikingly found that $T_c$ increases with pressure, reaching 25 K at 161 GPa, a record high among all known elemental superconductors.[9]

Extensive debates have been recently focused on the validity of the observed SC phase.[10-20] *Ab initio* phonon calculations within semi-local approximation (SLA) showed that the SC structure exhibits severely imaginary (unstable) phonons.[12, 13] Subsequently, structure prediction studies[10, 21] proposed a body-centered tetragonal $I4_1/amd$ structure, which is calculated to be energetically more stable than the SC structure within SLA over the entire pressure range of Ca-III. These calculations are apparently against the experimental observation of SC phase. To minimize these discrepancies, many theoretical investigations[14-16, 19, 21] proposed that anharmonic effects may help to stabilize the phonons of SC structure at room temperature. Indeed, Errea *et al.* have computed that the strong quantum anharmonic effects do stabilize the phonons of SC phase even at zero kelvin.[20] More recent quantum Monte Carlo calculations[19] demonstrated that SC structure at room temperature can be energetically more stable than $I4_1/amd$ structure at 50 GPa. These excellent theoretical contributions improved us the understanding of the discrepancy between theory and experiment.

In view of the contradiction between theory and experiment on SC phase, further experimental efforts have been greatly devoted to the structural solution of Ca-III. Mao *et al.*[17] observed a slight broadening of (111) Bragg peak and thus suggested an alternative rhombohedral structure of $R$-$3m$ at room temperature, which transforms to an orthorhombic $Cmmm$ structure on cooling to 30 K at 44 GPa.[7, 17] Instead, a very recent experiment confirmed the earlier observed SC structure over a broad pressure range 30 - 90 GPa at temperatures ranging from 300 K to 40 K.[14] The broadening of Bragg peaks was ascribed to the large amplitude fluxional behaviour of Ca. It is exciting that when temperature was lowered to ~14 K, a new Ca-III' phase was synthesized at the pressure regime of 40-45 GPa right before entering the SC phase.[14] Unfortunately, the structure of Ca-III' remains unsolved.

Since the large anharmonic effects in Ca was attributed to the nesting of the Fermi surfaces,[12, 13] it is important to re-examine the band structures and total energies with improved Kohn-Sham density functional theory (KS-DFT) methods. KS-DFT method has been widely applied to solving different problems over a range of complex materials. Although the calculated chemical and physical properties are often in good agreement with experiments, there are several shortcomings of the standard DFT approach. For example, KS-DFT calculations within SLA often underestimate the energy band gaps and in some cases, even fail on the general characteristic of the electronic band structures. This failure of DFT is largely attributed to the deficient SLA (e.g., local-density approximation or generalized gradient approximation), which is unable to correctly describe the electrons exchange interaction. Teweldeberhan et al.[19] have pointed out that standard DFT calculations within SLA are insufficient to accurately account for the high-pressure energetic stability of Ca. To overcome this problem, a solution is to add certain amount of non-local Hartree-Fock (HF) exchange to the semi-local density functional. This is the so-called PBE0 hybrid functional method,[22] which contains 25 % of the exact HF exchange, 75 % of the Perdew-Burke-Ernzerhof (PBE)[23] exchange, and 100 % of the PBE correlation energy. As an alternative, Heyd-Scuseria-Erhzerhof (HSE)[24] proposed a screened Coulomb hybrid functional by separating the short-range and long range HF exchange, which offers highly efficient computations on extended periodic systems. As exemplified, the HSE calculated energy difference between the diamond and $\beta$-tin phases of Si agrees best with the diffusion Monte Carlo results while other functionals (local-density approximation, Perdew-Wang 91, PBE, Tao-Perdew-Staroverov-Scuseria) largely underestimate the energy difference by about 140-250 meV/atom.[25]

calculations established that the long-thought $I4_1/amd$ structure remains stable in a narrow pressure range before entering SC phase, allowing us to assign the $I4_1/amd$ structure as the experimentally synthesized Ca-III' phase by further evidence of the excellent agreement between our simulated XRD patterns and the experimental data.

The underlying *ab initio* structural relaxations were performed in the framework of DFT within the PBE, HSE and PBE0 functionals, as implemented in the VASP (Vienna *ab initio* Simulation Package) code[26, 27]. The all-electron projector-augmented wave (PAW) method[28, 29] was adopted with the PAW potentials taken from the VASP library where 10 electrons ($3s^2 3p^6 4s^2$) were treated as valence electrons for Ca. A plane wave energy cutoff of 500 eV was employed. The Monkhorst-Pack $k$ grids[30] of 16×16×16 for BCC structure, 16×16×16 for SC structure, 10×10×16 for *Cmmm* structure, and 14×14×14 for $I4_1/amd$ structure, respectively, were used to ensure that all the enthalpy calculations are well converged to better than 1 meV/atom. Phonons were calculated with a supercell method as implemented in the PHONOPY program.[31] This method uses the Hellmann-Feynman forces calculated from the optimized supercell through VASP based on DFT within PBE and HSE functionals, respectively. Convergence check gave the use of a 4×4×4 supercell and 4×4×4 k-meshes in the Brillouin Zone (BZ).

TABLE I. Calculated lattice constants (*a*) of SC phase in Ca within PBE, HSE, and PBE0 functionals at 50 GPa in comparison with experiment.

|     | PBE   | HSE   | PBE0  | Experiment |
|-----|-------|-------|-------|------------|
| a/Å | 2.571 | 2.587 | 2.587 | 2.606[a]   |

[a]Reference[5]

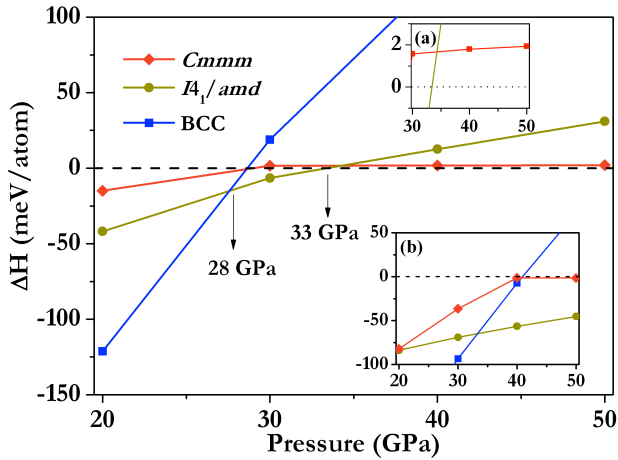

**Fig. 1 (color online)** Enthalpies (relative to SC structure) of *Cmmm*, $I4_1/amd$ and BCC structures within HSE (main figure and inset (a)) and PBE (inset (b)) functionals as a function of pressure. Our PBE0 enthalpy calculations at 30 and 50 GPa confirmed the results of HSE.

We here have performed more reliable DFT calculations within the HSE and PBE0 functionals to re-examine the total energies, electronic band structures, and phonons of BCC, SC, $I4_1/amd$, and *Cmmm* phases of Ca. It is fundamentally demonstrated that the SC phase is thermodynamically most stable among all known phases above 33 GPa and is dynamically stable without exhibiting any imaginary phonons, indicating SC is the optimum structure of Ca-III. Furthermore, our hybrid functional

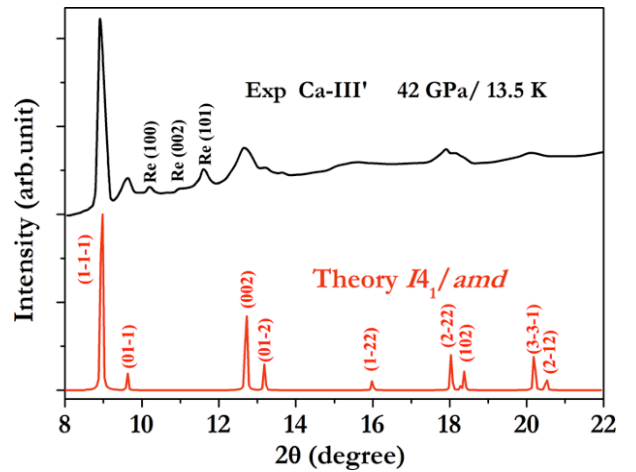

**Fig. 2 (color online)** The simulated XRD patterns of $I4_1/amd$ at 42 GPa along with the experimental XRD data of Ca- III' phase (Ref. 14). The lattice parameters of $I4_1/amd$ take a = b = 2.56 Å, c = 2.77 Å at 42 GPa.

In order to investigate the thermodynamic stability of Ca, the computed enthalpies of various structures in the pressure range 20-50 GPa within HSE and PBE functionals are shown in Fig. 1. Consistent with the experimental observation, the BCC structure is the most stable structure at 20-28 GPa (Fig. 1). Within PBE functional (inset (b) of Fig. 1), the $I4_1/amd$ structure becomes more stable than BCC structure at pressures above 33 GPa and SC (or $Cmmm$) structure is never stable in the entire pressure range studied. Our PBE results are in excellent agreement with previously theoretical calculations.[10, 11] However, once more precise HSE functional was considered, both SC and $Cmmm$ structures become energetically preferable over $I4_1/amd$ structure above 33 GPa (Fig.1). Note that SC structure has an almost constant lower enthalpy (~2 meV) than $Cmmm$ structure throughout the pressure range studied (inset of Fig. 1). Our HSE calculations are able to energetically account for the experimental observations[5, 7, 14, 17] of SC phase. Moreover, we have also calculated lattice parameters of SC structure at 50 GPa for comparison with experiment (Table I). The lattice constant derived from hybrid functionals (HSE and PBE0) is 2.587 Å, in a better agreement with experimental data of 2.606 Å with only a discrepancy about 0.7%.

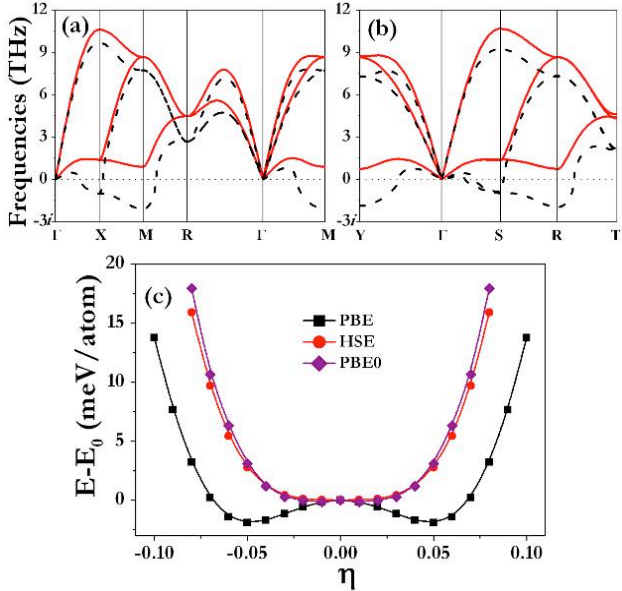

**Fig. 3 (color online)** Phonon dispersions of (a) SC and (b) $Cmmm$ structures at 50 GPa, respectively. (c) Total energies within PBE, HSE and PBE0 functionals for SC phase when the atoms are displaced along the eigenvector of transverse mode at $M$ point.

Very recently, John et al. observed the formation of a new Ca-III' structure at very low temperature (14 K) and pressures 40-45 GPa before Ca enters into the SC structure upon compression.[14] However, their obtained XRD data are not in sufficiently high quality to reliably index the structure of Ca-III'.[14] It is noteworthy that our enthalpy calculations (Fig. 1) within hybrid functional established the phase stability of SC structure, but did not exclude the theoretical $I4_1/amd$ structure. In fact, the $I4_1/amd$ structure still remains stable in a narrow pressure range (28-33 GPa) intermediated between BCC and SC structures (Fig. 1). This naturally leads us to correlate $I4_1/amd$ structure with Ca-III' phase though it is in apparently against to the argument of Ref. [14]. We have simulated XRD patterns of the $I4_1/amd$ structure to compare with the experimental XRD data on Ca-III' phase as shown in Fig. 2. It is remarkably found that the mutual agreement between theory and experiment is excellent (Fig. 2). It is therefore not unreasonable to propose the actual experimental synthesis of the $I4_1/amd$ structure predicted in 2009[21] and 2010[10], respectively, by John et al.[14]

Figure 3a presents the phonon spectra of the SC phase calculated within PBE and HSE functionals at 50 GPa. It is found that within PBE functional the SC phase has considerably imaginary phonons with the most severe imaginary phonons appearing at $M$ point, in excellent agreement with previously theoretical calculations.[12, 13, 19] Once the more precise HSE functional was applied to the phonon calculations, imaginary phonons vanish completely (Fig. 3a). This allows us to rationalize the structural stability of SC phase.

In order to probe the underlying mechanism on the failure of semi-local DFT theory on phonon calculations, we have simulated the total energies as a function of the atomic displacements along the eigenvector of the transverse phonon mode at M point for SC phase within different functional schemes of PBE, HSE and PBE0. For such a calculation, a supercell containing four Ca atoms, i.e., a doubling of the original cell along both x and y directions was constructed. Within PBE functional, it is seen that the potential energy associated with this distortion has a typical double-well character, which is in good agreement with the results of Refs. 12, 19. This double-well potential feature leads to the calculated imaginary phonons. In contrast, use of the hybrid functionals (HSE or PBE0) destroys completely the double-well potential and stabilizes a single-well potential (Fig. 3b). As a consequence, our hybrid functional calculations established the dynamical stability of SC structure. We point out that it is unnecessary to include the anharmonic effects[20] or thermal phonon corrections[14, 19] to stabilize the phonons, where a harmonic approximation is sufficient to account for the dynamical stability of SC phase.

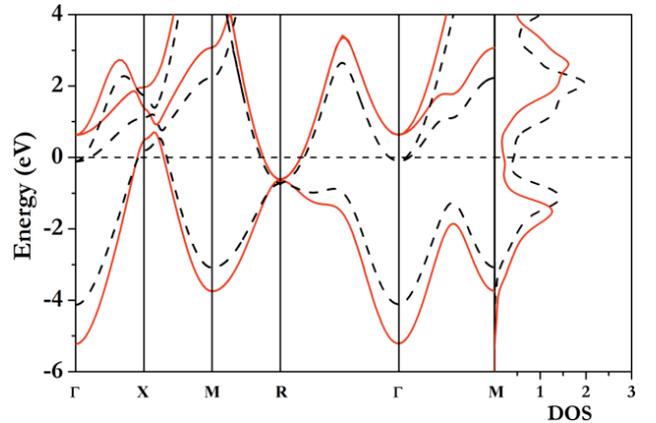

**Fig. 4 (color online)** Electronic band structures and DOS (in unit of $eV^{-1}$ per atom) of SC phase derived from PBE (red solid lines) and HSE (black dashed lines) functionals at 50 GPa, respectively. The Fermi energy was set to zero.

It is known that the screened hybrid functional of HSE has the ability to accurately predict electronic band structures.[24] We therefore have re-examined the band structures of SC phase as presented in Fig. 4. The semi-local DFT calculations revealed that three bands cross over the Fermi level, leading to two electron pockets at $R$ and $\Gamma$ points, and one electron hole at $X$ point. In contrast, HSE functional gives a band structure dramatically different with that of standard semi-local DFT. The conduction band was significantly lifted to higher energy, while the valence band was downshifted to lower energy. As a result, the electron pocket at $\Gamma$ point vanishes and a smaller electron density of states at Fermi Level was evidenced (Fig. 4), which might be the physical origin of lowering the total energy of SC phase in the hybrid functional calculations.

Previous experiment[17] observed the phase transformation of SC→ *Cmmm* on cooling to 30 K at 44 GPa. Note that the *Cmmm* structure (the lattice parameters in a primitive cell take $a = b = 2.5842$ Å, $c = 2.5843$ Å, $\alpha = \beta = 90°$ and $\gamma = 89.8°$ at 50 GPa) can be viewed as a slightly distorted SC structure. As a result, our enthalpy calculations (Fig. 1) predicted that the SC and *Cmmm* structures have very competitive enthalpies. Earlier DFT phonon calculations within SLA predicted that *Cmmm* structure is dynamically unstable,[19] in agreement with our PBE results (Fig. 3b). Again, our use of hybrid functionals fundamentally eliminated the imaginary phonons of *Cmmm* structure, supporting its structural stability. In order to energetically reproduce the temperature-induced SC→ *Cmmm* transition, we tried to examine the Gibbs free energies of SC and *Cmmm* structures by using quasi-harmonic approximation.[32] We found at such a low temperature 30 K, entropy term is not important but zero-point energy (ZPE) dominates the vibrational energy. At 50 GPa, our calculated ZPEs of *Cmmm* and SC structures are 34.40 and 34.44 meV/atom, respectively. However, the resultant ZPE difference between the two structures is extremely small (<0.1meV/atom). Therefore, even by actual consideration of the vibrational energy, our theory is not able to account for the observed SC→ *Cmmm* transition. It is noteworthy that the recent proposed large amplitude fluxional behaviour of Ca[14] is not included in our static calculations. This might cause some errors in the calculations of vibrational energy. Nevertheless, the recent powder experiment[14] failed to observe the formation of *Cmmm* structure. Note that the recent diffusion quantum Monte Carlo calculations correctly account for the SC→*Cmmm* structural transitions at lowering temperature; however, no stable pressure range of $I4_1/amd$ structure was predicted.[19] It is clear that more research efforts are still needed to the understanding of low temperature phase stability of Ca.

In summary, we have performed more reliable DFT calculations by adopting hybrid functionals (HSE and PBE0) for the electron exchange energy to investigate the phase stability of SC structure of Ca over a wide pressure range. Our calculations have fundamentally revised earlier theoretical understanding of SC phase based on SLA. We found that SC phase is energetically most stable among all known phases above 33 GPa and contains intrinsically no any imaginary phonons within the harmonic approximation. Our results provided a complete rationalization of the SC phase of Ca. Moreover, our hybrid functional calculations established the phase stability of $I4_1/amd$ structure in a very narrow pressure range 28-33 GPa, intermediating between BCC and SC structures. This leads us to correlate $I4_1/amd$ structure with the newly synthesized Ca-III' phase at 14 K.[14] Careful comparison between the simulated XRD patterns of $I4_1/amd$ structure and the experimental XRD data of Ca-III' phase allows us to assign the $I4_1/amd$ structure to Ca-III' phase. Our results highlight the important role played by the precise electron exchange energy in dealing with Ca and place considerable warnings on the researches of other materials, where standard SLA-DFT calculations are frequently applied to.

The authors acknowledge the funding supports from National Natural Science Foundation of China under Grant Nos. 11274136, 11025418 and 91022029, China 973 Program under Grant No. 2011CB808200, 2012 Changjiang Scholar Program of China, and Changjiang Scholar and Innovative Research Team in University (No. IRT1132). Part of calculations was performed in high performance computing center of Jilin University.

*Author to whom correspondence should be addressed: mym@jlu.edu.cn